\documentclass{JHEP3}
\usepackage{graphicx,amsmath,amssymb}
\usepackage{epsfig,multicol}
\usepackage{epsf}
\usepackage{epstopdf}
\input{epsf.sty}

\usepackage{graphicx}
\usepackage{amssymb}
\usepackage{amsmath}

\def\p{\partial}

\def\half{{1\over 2}}
\def\({\left(}
\def\){\right)}
\def\[{\left[}
\def\]{\right]}

\def\e{\begin{equation}}
\def\q{\end{equation}}
\def\m{\begin{eqnarray}}
\def\n{\end{eqnarray}}
\title{Scale dependences of local form non-Gaussianity parameters from a DBI isocurvature field}
\author{Qing-Guo Huang $^1$ \footnote{huangqg@itp.ac.cn} \ and Chunshan Lin $^2$ 
\\\small{\em
$^1$ Key Laboratory of Frontiers in Theoretical Physics, Institute
of Theoretical Physics, Chinese Academy of Sciences, Beijing
100190, China}
\\\small{\em
$^2$ Institute for the Physics and Mathematics of the Universe, The University of Tokyo, Kashiwa, 277-8583, Japan}}

\abstract{
We derive the spectral indices and their runnings of local form $f_{NL}$ and $g_{NL}$ from a  DBI isocurvature field and we find that the indices are suppressed by the sound speed $c_s$.  This effect can be interpreted by the Lorentz boost from the viewpoint in the frame where brane is moving. 
}


\begin{document}

\section{Introduction}

Non-Gaussianity \cite{Bartolo:2004if} has become a very important probe to the physics in the early universe. It is helpful to figure out the mechanism for generating primordial curvature perturbation. A well-defined non-Gaussianity is the so-called local form non-Gaussianity which says that the curvature perturbation can be expanded to the non-linear orders at the same spatial point
\m
\zeta({\bf x})=\zeta_g({\bf x})+{3\over 5}f_{NL}\zeta_g^2({\bf x})+{9\over 25}g_{NL}\zeta_g^3({\bf x})+\dddot \ ,
\label{ng}
\n
where $\zeta_g$ is the Gaussian part of curvature perturbation, $f_{NL}$ and $g_{NL}$ are the non-Gaussianity parameters which characterize the sizes of local form bispectrum and trispectrum respectively. Single-field inflation model predicts $f_{NL}\sim {\cal O}(n_s-1)\sim {\cal O}(10^{-2})$ \cite{Maldacena:2002vr}. A convincing detection of a large local form non-Gaussianity is the smoking gun for the multi-field inflation in which the quantum fluctuations of the isocurvature field contribute to the final curvature perturbation on the super-horizon scales. See, for example, \cite{Enqvist:2001zp,Lyth:2001nq,Moroi:2001ct,Sasaki:2006kq,Huang:2008ze,Sasaki:2008uc,Li:2008fma,Huang:2008rj,Huang:2008zj,Cai:2008if,Kawasaki:2008pa,Chingangbam:2009xi,Cai:2009hw,Huang:2009vk,Enqvist:2009zf,RenauxPetel:2009sj,Zhang:2009gw,Choi:2010re} etc.

In the literatures the non-Gaussianity parameters are taken as a constant. However, recently there are some hints at a possible scale dependence of $f_{NL}$ come from the numerous observations of massive high-redshift clusters \cite{Jee:2009nr,Stott:2010zr,Brodwin:2010ig,Santos:2011qs} which seems in greater abundances than expected from a Gaussian statistics \cite{Hoyle:2010ce,Chongchitnan:2011eq}. In fact, $g_{NL}$ may also be scale dependent. 
The scale dependences of $f_{NL}$ and $g_{NL}$ are measured by their spectral indices $n_{NL}$ and $n_{g_{NL}}$ and their runnings $\alpha_{f_{NL}}$ and $\alpha_{g_{NL}}$ which are respectively defined by 
\m
f_{NL}(k)&=&f_{NL}(k_p)\({k\over k_p}\)^{n_{f_{NL}}+\half \alpha_{f_{NL}}\ln {k\over k_p}},\\
g_{NL}(k)&=&g_{NL}(k_p)\({k\over k_p}\)^{n_{g_{NL}}+\half \alpha_{g_{NL}}\ln {k\over k_p}}, 
\n
where $k_p$ is a pivot scale. The authors in \cite{Sefusatti:2009xu} showed that 
Planck \cite{:2006uk} and CMBPol \cite{Baumann:2008aq} are able to provide a 1-$\sigma$ uncertainty on the spectral index of $f_{NL}$ as follows
\m
\Delta n_{f_{NL}}\simeq 0.1 {50\over f_{NL}}{1\over \sqrt{f_{sky}}}\quad \hbox{for Planck},
\n
and
\m
\Delta n_{f_{NL}}\simeq 0.05 {50\over f_{NL}}{1\over \sqrt{f_{sky}}}\quad \hbox{for CMBPol},
\n
where $f_{sky}$ is the sky fraction. The effects on the halo bias from the scale dependent $f_{NL}$ are discussed in \cite{Becker:2010hx,Shandera:2010ei}. The studies on fingerprints of the scale-dependent $g_{NL}$ in CMB and large-scale structure are called for in the near future.

Actually the scale-independent non-Gaussianity parameters are not generic predictions of inflation model. The large scale dependences of non-Gaussianity parameters can be obtained in the axion N-flation with different decay constants for different axion fields \cite{Huang:2010es} and in the model with self-interacting canonical isocurvature field \cite{Byrnes:2009pe,Byrnes:2010ft,Byrnes:2010xd,Huang:2010cy,Riotto:2010nh,Huang:2011di,Byrnes:2011gh}. It is interesting for us to ask how the results should be modified for the non-canonical isocurvature field. In particular, we are curious that the scale dependence is enhanced or suppressed by the sound speed. For simplicity, we focus on the case with a DBI isocurvature field.

Our paper is organized as follows. In Sec.~2 we will discuss the dynamics and quantum fluctuation of a DBI isocurvature field. In Sec.~3 the spectral indices of $f_{NL}$ and $g_{NL}$ from a DBI isocurvature field will be calculated. Some discussions are contained in Sec.~4.

\section{Dynamics and quantum fluctuation of a DBI isocurvature field}

Brane inflation \cite{Dvali:1998pa} is considered to be a popular inflation model embedded into string theory. A more realistic setup is that a D3-brane moves in the internal six-dimensional Calabi-Yau manifold which contains one or more throats in Type IIB string theory \cite{Kachru:2003aw}. For simplicity, we consider a D3-brane which is mobile in a throat whose metric is given by 
\begin{eqnarray}
ds^2=h^{-1/2}g_{\mu\nu}dx^{\mu}dx^{\nu}+h^{1/2}\left[dr^2+b^2(r)d\theta^2+\dddot \ \right]~,
\end{eqnarray}
Here $\theta$ is an angular coordinate which is transverse to the radius direction $r$ and $b(r)$ is the radius of throat at $r$. In this paper, we choose signature $(-,+,+,+)$. The action for the mobile D3-brane is given by 
\m
S=-T_3\int d^4x \[\sqrt{-{\rm det}\[h^{-1/2} (g_{\mu\nu}+h\p_\mu r\p_\nu r+hb^2 \p_\mu\theta \p_\nu\theta)\]}-h^{-1}\]-\int d^4x \sqrt{-g} V(r,\theta). 
\n
It is convenient to define two canonical fields in the slow-rolling limit as follows
\m
\phi&=&\sqrt{T_3}r,\\
\sigma&=&\sqrt{T_3}b\theta.
\n
Here we consider $\dot \sigma\ll \dot \phi$, and then $\phi$ and $\sigma$ are taken to be the adiabatic and entropic direction during inflation respectively. 

In the limit of $f{\dot \phi}^2\ll 1$ and $f{\dot \sigma}^2\ll 1$, these two-field inflation is reduced to the canonical case, where $f=h/T_3$. In this paper we focus on another limit in which 
\m
c_s\simeq \sqrt{1-f{\dot \phi}^2}\ll 1, 
\n 
and 
\m
{\dot \sigma}^2/{\dot \phi}^2\ll c_s^2. 
\n
From the above action the equations of motion for $\phi$ and $\sigma$ are roughly given by 
\m
\ddot \phi+3H(1-\kappa/3)\dot \phi &\simeq& {1\over 2f^2}{\p f\over \p \phi} -c_s{\p V(\phi,\sigma)\over \p \phi},\\
3H \dot \sigma &\simeq& c_s \(3{\eta_b\over c_s}H^2\sigma-{\p V(\phi,\sigma)\over \p\sigma}\), \label{ds}
\n
where
\m
\kappa&\equiv& {\dot c_s\over H c_s}, \\
\eta_b&\equiv& {\dot b\over Hb},
\n
and both $\kappa$ and $\eta_b$ are assumed to be much smaller than unity. 
For simplicity, the cross-coupling between $\phi$ and $\sigma$ is assumed to be negligibly small, and Eq.~(\ref{ds}) is valid when 
\m
\left| {V''(\sigma)\over 3H^2} \right| \ll c_s^{-1}. 
\n
This is the slow-roll condition for the isocurvature field $\sigma$.
Now the dynamics of $\sigma$ becomes 
\m
3H \dot \sigma \simeq -c_s \[{\tilde m}^2\sigma+V'(\sigma)\],
\label{eom}
\n
where 
\m
{\tilde m}^2\equiv -3\eta_bH^2/c_s. 
\n
The slow variation of the radius of throat induces an effective mass for the isocurvature field along the angular direction.

The quantum fluctuations of $\phi$ and $\sigma$ have been well studied in \cite{Langlois:2008wt}. See also \cite{RenauxPetel:2009sj,Mizuno:2009cv,Gao:2009at}. Here we don't want to repeat the computations. We will only briefly recall the results in \cite{Langlois:2008wt}. The canonically normalized quantum fluctuations of $\phi$ and $\sigma$ are respectively given by 
\m
v_\phi={a\over c_s^{3/2}} \delta \phi,\quad v_\sigma={a\over \sqrt{c_s}} \delta\sigma, 
\n
whose Fourier modes corresponding to the Minkowski-like vacuum on very small scales are given by 
\m
v_{\phi,k}\simeq v_{\sigma,k}\simeq {1\over \sqrt{2kc_s}}e^{-ikc_s\chi}\(1-{i\over k c_s \chi}\),
\n
where 
\m
\chi=\int {dt \over a(t)}
\n
is the conformal time. The perturbation mode of $k$ exits horizon at the time of $c_sk=aH$. Therefore the power spectra for $\delta \phi$ and $\delta \sigma$ are 
\m
P_{\delta \phi}&\simeq& \({H\over 2\pi}\)^2,\\
P_{\delta \sigma}&\simeq& \({H\over 2\pi c_s}\)^2. 
\n
The amplitude of $\delta \sigma$ is amplified by a factor of $1/c_s$ compared to the canonical one.

Before closing this section, we want to estimate the typical value of $\sigma$ during inflation. The quantity of $\langle \sigma^2 \rangle$ coming from its quantum fluctuations is 
\m
\langle \sigma^2 \rangle={1\over (2\pi)^3}\int |\delta \sigma_k|^2 d^3k= {1\over (2\pi)^3} \int \({1\over 2a^2k}+{H^2\over 2k^3c_s^2}\)d^3k. 
\n
Considering that the physical momentum $p$ is related to $k$ by $p=k/a=e^{-Ht}k$, we have
\m
\langle \sigma^2 \rangle={1\over (2\pi)^3}\int {d^3p\over p} \(\half + {H^2\over 2p^2 c_s^2}\).
\n
The first term is contributed from vacuum fluctuations in Minkowski space and it can be eliminated by renormalization. In addition, the physical momenta for the modes of quantum fluctuations we concern in the time interval between $0$ and $t$ are those from $c_s^{-1}H$ to $c_s^{-1}H e^{-Ht}$. Therefore 
\m
\langle \sigma^2 \rangle={H^2\over 4\pi^2 c_s^2}\int_{c_s^{-1}H e^{-Ht}}^{c_s^{-1}H} {dp\over p}={H^3\over 4\pi^2 c_s^2}t.
\n
It indicates that the quantum fluctuation of the field $\sigma$ in the inflation epoch can be modeled by a random walk with step size $H/2\pi c_s$ per Hubble time. However the vacuum expectation value of $\sigma$ cannot go like $t$ for $t\rightarrow \infty$ if $\sigma$ has a potential. For example, let's consider the potential of $\sigma$ is 
\m
V(\sigma)=\half m_\sigma^2 \sigma^2 + \lambda \sigma^n.
\n
From Eq.~(\ref{eom}), the dynamics of $\sigma$ is described by  
\m
3H\dot \sigma=-c_s m^2\sigma(1+n\lambda \sigma^{n-2}/m^2), 
\n
where 
\m
m^2=m_\sigma^2+{\tilde m}^2.
\n
Similar to \cite{Starobinsky:1994bd}, combining the contributions from quantum fluctuation and classical equation of motion, we have 
\m
{d\langle \sigma^2 \rangle \over dt}={H^3\over 4\pi^2c_s^2}-c_s{2m^2\over 3H}\langle \sigma^2\rangle \[1+{n\lambda\over m^2}\langle \sigma^2\rangle^{n/2-1}\]. 
\n
The above differential equation approaches a constant equilibrium value, namely
\m
\langle \sigma^2 \rangle={3H^4\over 8\pi^2m^2c_s^3}{1\over 1+{n\over 2}s },
\n
where
\m
s\equiv 2\lambda \langle \sigma^2\rangle^{n/2-1}/m^2. 
\n
The typical value of $\sigma$ during inflation is $\sigma_*=\sqrt{\langle \sigma^2\rangle}\sim c_s^{-3/2}$ which is enhanced for $c_s\ll 1$.

\section{Scale dependence of local form non-Gaussianity parameters}

In this paper we consider that the curvature perturbation is generated by the isocurvature field $\sigma$ at the end of inflation or deep in the radiation dominated era, and the curvature perturbation can be expanded to the non-linear orders by using the $\delta N$ formalism \cite{Starobinsky:1986fxa}: 
\m
\zeta(t_f,{\bf x})=N_{,\sigma}(t_f,t_i)\delta \sigma(t_i,{\bf x})+\half N_{,\sigma\sigma}\delta\sigma^2(t_i,{\bf x})+{1\over 6} N_{,\sigma\sigma\sigma}\delta\sigma^3(t_i,{\bf x})+\dddot \ ,
\label{deltan}
\n
where $N_{,\sigma}$, $N_{,\sigma\sigma}$ and $N_{,\sigma\sigma\sigma}$ are the first, second and third order derivatives of the number of e-folds with respect to $\sigma$ respectively. Here $t_f$ denotes a  final uniform energy density hypersurface and $t_i$ labels any spatially flat hypersurface after the horizon exit of a given mode. Similar to \cite{Huang:2010cy,Huang:2011di}, $t_i$ is set to be $t_*(k)$ which is determined by $k=a(t_*)H_*$ for a given mode with comoving wavenumber $k$. $H_*$ denotes the Hubble parameter during inflation from now on. 

From Eq.~(\ref{deltan}), the amplitude of curvature perturbation is given by
\e
\Delta_{\cal R}^2=N_{,\sigma}^2(t_*)\(H_*\over 2\pi c_s\)^2.
\label{pnh}
\q
The amplitude of Gravitational waves perturbation only depends on the energy scale of inflation as follows 
\m
\Delta_T^2={H_*^2\over \pi^2/2}.
\n
Here we work on the unit of $M_p=1$. The scale dependence of gravitational wave perturbation is measured by $n_T$ which is defined by
\m
n_T\equiv {d\Delta_T^2\over d\ln k}=-2\epsilon_H,
\n
where
\m
\epsilon_H&\equiv&-{\dot H_*\over H_*^2}.
\n
For convenience, the tensor-scalar ratio $r_T$ is introduced to measure the amplitude of gravitational waves:
\e
r_T\equiv \Delta_T^2/\Delta_{\cal R}^2={8c_s^2\over N_{,\sigma}^2(t_*)}.
\q
Comparing Eq.~\eqref{deltan} to (\ref{ng}), the non-Gaussianity parameters are given by 
\e
f_{NL}={5\over 6}{N_{,\sigma\sigma}(t_*)\over N_{,\sigma}^2(t_*)}.
\q
and
\m
g_{NL}={25\over 54}{N_{,\sigma\sigma\sigma}(t_*)\over N_{,\sigma}^3(t_*)}.
\n

Following the method in \cite{Huang:2010cy,Huang:2011di}, we introduce a new time $t_r (>t_*)$ which is chosen as a time soon after all the modes of interest exit the horizon during inflation and keep it fixed.  The value of $\sigma$ at $t_r$ is related to that at time $t_*$ and the time $t_*$ by
\e
\int_{\sigma_*}^{\sigma_r}{d\sigma\over V_{\rm eff}'(\sigma)}=-\int_{t_*}^{t_r} {c_s(t)\over 3H(t)} dt, 
\q
where 
\m
V_{\rm eff}=\half {\tilde m}^2\sigma^2+V(\sigma). 
\n
The equation of motion in Eq.~(\ref{eom}) can be rewritten by
\m
3H\dot \sigma=-c_sV_{\rm eff}'.
\label{neom}
\n
Therefore we have
\m
\left. {\p \sigma_r\over \p \sigma_*} \right|_{t_*}&=&{V_{\rm eff}'(\sigma_r)\over V_{\rm eff}'(\sigma_*)}, \\
\left. {\p \sigma_r\over \p t_*} \right|_{\sigma_*}&=&{c_s(t_*) V_{\rm eff}'(\sigma_r)\over 3H(t_*)}.
\n
Considering
\m
{d\over dt_*}F(\sigma_r)={\p F(\sigma_r)\over \p\sigma_r}(\dot \sigma_* {\p \sigma_r\over \p \sigma_*}+{\p \sigma_r\over \p t_*})
\n
and Eq.~(\ref{neom}), one finds 
\e
{d\over d\ln k}F(\sigma_r)={d\over H_*dt_*}F(\sigma_r)=0,
\q
which implies that $F(\sigma_r)$ is scale independent. \footnote{More precisely, the perturbation mode with $k$ exits horizon when $c_sk=a_*H_*$, and then $d\ln k=(1-\epsilon_H-\kappa)H_*dt_*\simeq H_*dt_*$. } 
Taking into account that $\sigma_r$ is a function of $\sigma_*$, we have
\m
N_{,\sigma}(t_*)&=&{\p\sigma_r\over \p\sigma_*} {\p N(\sigma_r)\over \p\sigma_r},\\
N_{,\sigma\sigma}(t_*)&=&{\p^2\sigma_r\over \p\sigma_*^2} {\p N(\sigma_r)\over \p\sigma_r}+\({\p\sigma_r\over \p\sigma_*}\)^2  {\p^2 N(\sigma_r)\over \p\sigma_r^2},\\
N_{,\sigma\sigma\sigma}(t_*)&=&{\p^3\sigma_r\over \p\sigma_*^3} {\p N(\sigma_r)\over \p\sigma_r}+3{\p\sigma_r\over \p\sigma_*}{\p^2\sigma_r\over \p\sigma_*^2} {\p^2 N(\sigma_r)\over \p\sigma_r^2}+\({\p\sigma_r\over \p\sigma_*}\)^3  {\p^3 N(\sigma_r)\over \p\sigma_r^3}.
\n
Since ${\p N(\sigma_r)\over \p\sigma_r}$, ${\p^2 N(\sigma_r)\over \p\sigma_r^2}$ and ${\p^3 N(\sigma_r)\over \p\sigma_r^3}$ are scale independent, one obtains
\m
{d\ln N_{,\sigma}(t_*)\over d\ln k}&=&c_s\eta_{\sigma\sigma},\\
{d\ln N_{,\sigma\sigma}(t_*)\over d\ln k}&=&2c_s\eta_{\sigma\sigma}+c_s \eta_3 {N_{,\sigma}(t_*)\over N_{,\sigma\sigma}(t_*)},\\
{d\ln N_{,\sigma\sigma\sigma}(t_*)\over d\ln k}&=&3c_s\eta_{\sigma\sigma}+3c_s\eta_3 {N_{,\sigma\sigma}(t_*) \over N_{,\sigma\sigma\sigma}(t_*)}+c_s \xi_4 {N_{,\sigma}(t_*) \over N_{,\sigma\sigma\sigma}(t_*)},
\n
where the slow-roll equation of motion for $\sigma$ is considered and
\m
\eta_{\sigma\sigma}\equiv {V_{\rm eff}''(\sigma_*)\over 3H_*^2},\quad
\eta_3\equiv {V_{\rm eff}'''(\sigma_*)\over 3H_*^2},\quad
\xi_4={V_{\rm eff}^{(4)}(\sigma_*)\over 3H_*^2}.
\n
From the above results, the spectral indices of $\Delta_{\cal R}^2$, $f_{NL}$ and $g_{NL}$ are respectively given by
\m
n_s\equiv 1+{d\ln \Delta_{\cal R}^2\over d\ln k}=1-2\epsilon_H-2\kappa+2c_s\eta_{\sigma\sigma}, 
\label{ns}
\n
\m
n_{f_{NL}}\equiv {d\ln |f_{NL}|\over d\ln k}=c_s \eta_3 {N_{,\sigma}(t_*)\over N_{,\sigma\sigma}(t_*)},
\n
and
\m
n_{g_{NL}}&\equiv& {d\ln |g_{NL}|\over d\ln k}=3c_s\eta_3 {N_{,\sigma\sigma}(t_*)\over N_{,\sigma\sigma\sigma}(t_*)}+c_s\xi_4 {N_{,\sigma}(t_*)\over N_{,\sigma\sigma\sigma}(t_*)}. 
\n
We see that the spectral indices of both $f_{NL}$ and $g_{NL}$ are suppressed by a factor of $c_s$ for $c_s\ll 1$. 


Similar to \cite{Huang:2010cy,Huang:2011di}, the running of spectral indices of $f_{NL}$ and $g_{NL}$ are defined by 
\m
\alpha_{f_{NL}}\equiv {dn_{f_{NL}}\over d\ln k}&=& (\kappa+2\epsilon_H-c_s\eta_{\sigma\sigma}-c_s \eta_4)n_{f_{NL}}-n_{f_{NL}}^2, \\
\alpha_{g_{NL}}\equiv  {dn_{g_{NL}}\over d\ln k}&=& (\kappa+2\epsilon_H-2c_s\eta_{\sigma\sigma})n_{g_{NL}}-n_{g_{NL}}^2 \nonumber \\
&+&{1\over g_{NL}}\[2f_{NL}^2(c_s\eta_{\sigma\sigma}-c_s\eta_4+n_{f_{NL}})n_{f_{NL}}-{25\over 432}\xi_4\xi_5 r_T\],
\n
where
\m
\eta_4={V_{\rm eff}'V_{\rm eff}^{(4)}\over 3H^2V'''},\quad 
\xi_5={V_{\rm eff}'V_{\rm eff}^{(5)}\over 3H^2V_{\rm eff}^{(4)}}.
\n
The spectral indices of $f_{NL}$ and $g_{NL}$ are nice parameters to characterize the scale dependences of these two non-Gaussianity parameters only when $n_{f_{NL}}$ and $n_{g_{NL}}$ are much less than unity.

\subsection{DBI isocurvature field with polynomial potential}

In this subsection we will consider a simple example in which the DBI isocurvature field has an effective  polynomial potential 
\m
V_{\rm eff}(\sigma)=\half m^2\sigma^2 +\lambda \sigma^n, 
\n
and then  
\m
\eta_{\sigma\sigma}&=&{m^2\over 3H_*^2}\[1+n(n-1){s\over 2}\],\\
\eta_3&=&{\eta_{\sigma\sigma}\over \sigma_*}{n(n-1)(n-2)s/2 \over 1+n(n-1)s/2}. 
\n
Combing the normalization of curvature perturbation, we find 
\m
n_{f_{NL}} f_{NL}={\rm sign}(N_{,\sigma}){5\over 6 \Delta_{\cal R}} (c_s\eta_{\sigma\sigma}){H_*/2\pi c_s\over\sigma_*}{n(n-1)(n-2)s/2 \over 1+n(n-1)s/2},
\n
Here $\Delta_{\cal R}$ is normalized to be $4.96\times 10^{-5}$ by WMAP in \cite{Komatsu:2010fb}. The slow-roll condition for $\sigma$ is $c_s\eta_{\sigma\sigma}\ll 1$. $H_*/2\pi c_s$ is the amplitude of quantum fluctuation of $\sigma$ and it should be less than $\sigma_*$.  
For $c_s\eta_{\sigma\sigma}\sim 10^{-2}$, ${H_*/2\pi c_s\over \sigma_*}\sim 10^{-1}$ and $s\gtrsim 10^{-1}$, we have $|n_{f_{NL}} f_{NL}|\sim {\cal O}(10)$ which is detectable by PLANCK. In this case, if $c_s\lesssim 10^{-2}$, $\eta_{\sigma\sigma}\gtrsim 1$ which implies that the effective mass of $\sigma$ is not less than the Hubble parameter during inflation.

\begin{figure}[h]
\begin{center}
\includegraphics[width=12cm]{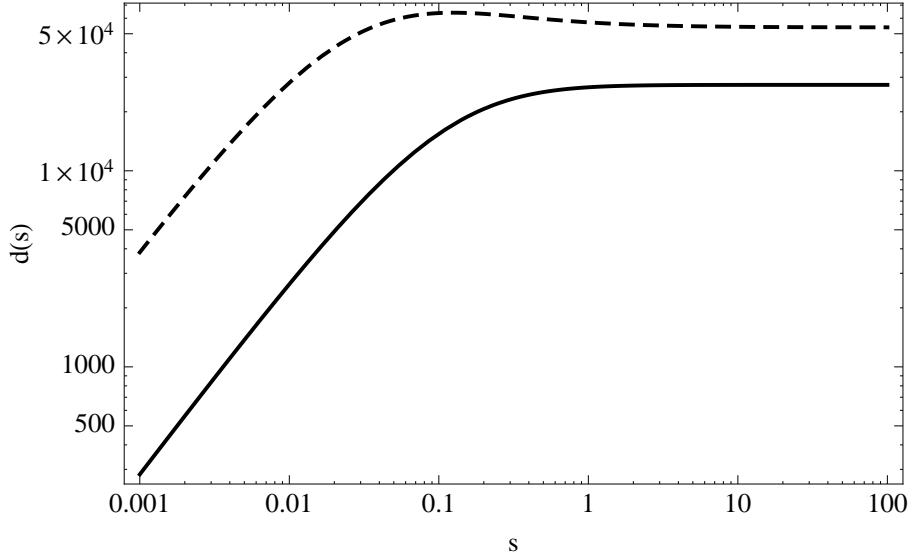}
\end{center}
\caption{The function of $d(s)$. The solid and dashed curves correspond to $n=4$ and $n=8$ respectively.  }
\label{fig:ds}
\end{figure}

Taking into account the typical value of $\sigma$ in Sec.~2, the above equation becomes
\m
n_{f_{NL}} f_{NL}={\rm sign}(N_{,\sigma}) (c_s \eta_{\sigma\sigma})^{3/2}d(s),  
\n
where 
\m
d(s)={5\sqrt{2}\over 6 \Delta_{\cal R}}\sqrt{1+ns/2\over 1+n(n-1)s/2}{n(n-1)(n-2)s/2 \over 1+n(n-1)s/2}
\n
which is illustrated in Fig.~\ref{fig:ds}. 
For $n=4$ and $s\gg 1$, $d(s)\simeq 2.7\times 10^4$, the scale dependence of $f_{NL}$ can be detected by PLANCK if $c_s\eta_{\sigma\sigma}\gtrsim 0.003$.

\section{Discussions}

In this paper we calculate the spectral indices of $f_{NL}$ and $g_{NL}$ and their runnings generated by a DBI isocurvature field on the super-horizon scales. We find that the indices are suppressed by the speed of sound $c_s$. This suppression effect is not surprised and one can understand it from the effect of Lorentz boost. Here $1/c_s$ is nothing but the Lorentz boost factor. The time interval in the frame where brane is moving is dilated by a factor $1/c_s$ compared to that in the frame where brane is at rest. Because the coordinate $\sigma$ is transverse to the motion direction of brane, the value of $\sigma$ does not change. Therefore the velocity of $\sigma$ is suppressed by a factor $c_s$. That is why there is a factor $c_s$ on the right hand side of Eq.~(\ref{eom}). On the other hand, the scale dependences of the non-Gaussianity parameters come from the small variation of isocurvature field $\sigma$ from $t_k$ to $t_{k+dk}$, where $t_k$ and $t_{k+dk}$ correspond to the time when perturbation modes $k$ and $k+dk$ exit horizon during inflation respectively. Therefore the variation of isocurvature field $\sigma$ is suppressed by $c_s$, and hence the spectral indices of $f_{NL}$ and $g_{NL}$ are suppressed by $c_s$ as well.

Even though the scale dependences of non-Gaussianity parameters generated by the DBI isocurvature field are probably detectable, its effective of mass is required to be comparable or larger than the Hubble parameter during inflation if $c_s\ll 1$.

There are so many models which can produce a large local form non-Gaussianity and an accurate measurement of $f_{NL}$ cannot help us to distinguish them. But once the scale dependences of the non-Gaussianity parameters are detected, we can get more information about the isocurvature field, such as how it interacts with itself.

\vspace{1.5cm}

\noindent {\bf Acknowledgments}

\vspace{.5cm}

QGH is supported by the project of Knowledge Innovation Program of
Chinese Academy of Science and a grant from NSFC.


\newpage

\begin{thebibliography}{99}




\bibitem{Bartolo:2004if}
  N.~Bartolo, E.~Komatsu, S.~Matarrese and A.~Riotto,
  ``Non-Gaussianity from inflation: Theory and observations,''
  Phys.\ Rept.\  {\bf 402}, 103 (2004)
  [arXiv:astro-ph/0406398].

\bibitem{Maldacena:2002vr}
  J.~M.~Maldacena,
  ``Non-Gaussian features of primordial fluctuations in single field inflationary models,''
  JHEP {\bf 0305}, 013 (2003).
  [astro-ph/0210603].
  


\bibitem{Enqvist:2001zp}
  K.~Enqvist and M.~S.~Sloth,
  ``Adiabatic CMB perturbations in pre big bang string cosmology,''
  Nucl.\ Phys.\  B {\bf 626}, 395 (2002)
  [arXiv:hep-ph/0109214].

\bibitem{Lyth:2001nq}
  D.~H.~Lyth and D.~Wands,
  ``Generating the curvature perturbation without an inflaton,''
  Phys.\ Lett.\  B {\bf 524}, 5 (2002)
  [arXiv:hep-ph/0110002].

\bibitem{Moroi:2001ct}
  T.~Moroi and T.~Takahashi,
  ``Effects of cosmological moduli fields on cosmic microwave background,''
  Phys.\ Lett.\  B {\bf 522}, 215 (2001)
  [Erratum-ibid.\  B {\bf 539}, 303 (2002)]
  [arXiv:hep-ph/0110096].  

\bibitem{Sasaki:2006kq}
  M.~Sasaki, J.~Valiviita and D.~Wands,
  ``Non-gaussianity of the primordial perturbation in the curvaton model,''
  Phys.\ Rev.\  D {\bf 74}, 103003 (2006)
  [arXiv:astro-ph/0607627].

\bibitem{Huang:2008ze}
  Q.~G.~Huang,
  ``Large Non-Gaussianity Implication for Curvaton Scenario,''
  Phys.\ Lett.\  B {\bf 669}, 260 (2008)
  [arXiv:0801.0467 [hep-th]].

\bibitem{Sasaki:2008uc}
  M.~Sasaki,
  ``Multi-brid inflation and non-Gaussianity,''
  Prog.\ Theor.\ Phys.\  {\bf 120}, 159 (2008)
  [arXiv:0805.0974 [astro-ph]].

\bibitem{Li:2008fma}
  S.~Li , Y.~F.~Cai and Y.~S.~Piao,
  ``DBI-Curvaton,''
  Phys.\ Lett.\  B {\bf 671}, 423 (2009)
  [arXiv:0806.2363 [hep-ph]].

\bibitem{Huang:2008rj}
  Q.~G.~Huang,
  ``The N-vaton,''
  JCAP {\bf 0809}, 017 (2008)
  [arXiv:0807.1567 [hep-th]].
  
\bibitem{Huang:2008zj}
  Q.~G.~Huang,
  ``A Curvaton with Polynomial Potential,''
  JCAP {\bf 0811}, 005 (2008)
  [arXiv:0808.1793 [hep-th]].

\bibitem{Cai:2008if}
  Y.~F.~Cai and W.~Xue,
  ``N-flation from multiple DBI type actions,''
  Phys.\ Lett.\  B {\bf 680}, 395 (2009)
  [arXiv:0809.4134 [hep-th]].
  
\bibitem{Kawasaki:2008pa}
  M.~Kawasaki, K.~Nakayama, T.~Sekiguchi, T.~Suyama and F.~Takahashi,
  ``A General Analysis of Non-Gaussianity from Isocurvature Perturbations,''
  JCAP {\bf 0901}, 042 (2009)
  [arXiv:0810.0208 [astro-ph]].    
    
\bibitem{Chingangbam:2009xi}
  P.~Chingangbam and Q.~G.~Huang,
  ``The Curvature Perturbation in the Axion-type Curvaton Model,''
  JCAP {\bf 0904}, 031 (2009)
  [arXiv:0902.2619 [astro-ph.CO]].

\bibitem{Cai:2009hw}
  Y.~-F.~Cai, H.~-Y.~Xia,
  ``Inflation with multiple sound speeds: a model of multiple DBI type actions and non-Gaussianities,''
  Phys.\ Lett.\  {\bf B677}, 226-234 (2009).
  [arXiv:0904.0062 [hep-th]].
  
  
\bibitem{Huang:2009vk}
  Q.~G.~Huang,
  ``A geometric description of the non-Gaussianity generated at the end of
  multi-field inflation,''
  JCAP {\bf 0906}, 035 (2009)
  [arXiv:0904.2649 [hep-th]].  

\bibitem{Enqvist:2009zf}
  K.~Enqvist, S.~Nurmi, G.~Rigopoulos, O.~Taanila and T.~Takahashi,
  ``The Subdominant Curvaton,''
  JCAP {\bf 0911}, 003 (2009)
  [arXiv:0906.3126 [astro-ph.CO]].


\bibitem{RenauxPetel:2009sj}
  S.~Renaux-Petel,
  ``Combined local and equilateral non-Gaussianities from multifield DBI
  inflation,''
  JCAP {\bf 0910}, 012 (2009)
  [arXiv:0907.2476 [hep-th]]. 

\bibitem{Zhang:2009gw}
  J.~Zhang, Y.~F.~Cai and Y.~S.~Piao,
  ``Rotating a Curvaton Brane in a Warped Throat,''
  JCAP {\bf 1005}, 001 (2010)
  [arXiv:0912.0791 [hep-th]].

\bibitem{Choi:2010re}
  K.~Y.~Choi and O.~Seto,
  ``Non-Gaussianity and gravitational wave background in curvaton with a double
  well potential,''
  arXiv:1008.0079 [astro-ph.CO].
  
   
  
  
\bibitem{Jee:2009nr}
  M.~J.~Jee {\it et al.},
  ``Hubble Space Telescope Weak-lensing Study of the Galaxy Cluster XMMU
  J2235.3-2557 at z=1.4: A Surprisingly Massive Galaxy Cluster when the
  Universe is One-third of its Current Age,''
  Astrophys.\ J.\  {\bf 704}, 672 (2009)
  [arXiv:0908.3897 [astro-ph.CO]].  
  
\bibitem{Stott:2010zr}
  J.~P.~Stott {\it et al.},
  ``The XMM Cluster Survey: The build up of stellar mass in Brightest Cluster
  Galaxies at high redshift,''
  Astrophys.\ J.\  {\bf 718}, 23 (2010)
  [arXiv:1005.4681 [astro-ph.CO]].


\bibitem{Brodwin:2010ig}
  M.~Brodwin {\it et al.},
  ``SPT-CL J0546-5345: A Massive z > 1 Galaxy Cluster Selected Via the
  Sunyaev-Zel'dovich Effect with the South Pole Telescope,''
  Astrophys.\ J.\  {\bf 721}, 90 (2010)
  [arXiv:1006.5639 [astro-ph.CO]].

\bibitem{Santos:2011qs}
  J.~S.~Santos, R.~Fassbender, A.~Nastasi, H.~Bohringer, P.~Rosati, R.~Suhada, D.~Pierini, M.~Nonino {\it et al.},
  ``Discovery of a massive X-ray luminous galaxy cluster at z=1.579,''
  [arXiv:1105.5877 [astro-ph.CO]].  


\bibitem{Hoyle:2010ce}
  B.~Hoyle, R.~Jimenez, L.~Verde,
  ``Implications of multiple high-redshift galaxy clusters,''
  Phys.\ Rev.\  {\bf D83}, 103502 (2011).
  [arXiv:1009.3884 [astro-ph.CO]].

\bibitem{Chongchitnan:2011eq}
  S.~Chongchitnan, J.~Silk,
  ``Primordial Non-Gaussianity and Extreme-Value Statistics of Galaxy Clusters,''
  [arXiv:1107.5617 [astro-ph.CO]].

\bibitem{Sefusatti:2009xu}
  E.~Sefusatti, M.~Liguori, A.~P.~S.~Yadav, M.~G.~Jackson and E.~Pajer,
  ``Constraining Running Non-Gaussianity,''
  JCAP {\bf 0912}, 022 (2009)
  [arXiv:0906.0232 [astro-ph.CO]].

\bibitem{:2006uk}
    [Planck Collaboration],
  ``Planck: The scientific programme,''
  arXiv:astro-ph/0604069.

\bibitem{Baumann:2008aq}
  D.~Baumann {\it et al.}  [CMBPol Study Team Collaboration],
  ``CMBPol Mission Concept Study: Probing Inflation with CMB Polarization,''
  AIP Conf.\ Proc.\  {\bf 1141}, 10 (2009)
  [arXiv:0811.3919 [astro-ph]].


\bibitem{Becker:2010hx}
  A.~Becker, D.~Huterer and K.~Kadota,
  ``Scale-Dependent Non-Gaussianity as a Generalization of the Local Model,''
  JCAP {\bf 1101}, 006 (2011)
  [arXiv:1009.4189 [astro-ph.CO]].

\bibitem{Shandera:2010ei}
  S.~Shandera, N.~Dalal and D.~Huterer,
  ``A generalized local ansatz and its effect on halo bias,''
  JCAP {\bf 1103}, 017 (2011)
  [arXiv:1010.3722 [astro-ph.CO]].


\bibitem{Huang:2010es}
  Q.~-G.~Huang,
  ``Scale dependence of $f_{NL}$ in N-flation,''
  JCAP {\bf 1012}, 017 (2010).
  [arXiv:1009.3326 [astro-ph.CO]].  



\bibitem{Byrnes:2009pe}
  C.~T.~Byrnes, S.~Nurmi, G.~Tasinato, D.~Wands,
  ``Scale dependence of local $f_NL$,''
  JCAP {\bf 1002}, 034 (2010).
  [arXiv:0911.2780 [astro-ph.CO]].
  
\bibitem{Byrnes:2010ft}
  C.~T.~Byrnes, M.~Gerstenlauer, S.~Nurmi, G.~Tasinato, D.~Wands,
  ``Scale-dependent non-Gaussianity probes inflationary physics,''
  JCAP {\bf 1010}, 004 (2010).
  [arXiv:1007.4277 [astro-ph.CO]].  

\bibitem{Byrnes:2010xd}
  C.~T.~Byrnes, K.~Enqvist, T.~Takahashi,
  ``Scale-dependence of Non-Gaussianity in the Curvaton Model,''
  JCAP {\bf 1009}, 026 (2010).
  [arXiv:1007.5148 [astro-ph.CO]].

\bibitem{Huang:2010cy}
  Q.~-G.~Huang,
  ``Negative spectral index of $f_{NL}$ in the axion-type curvaton model,''
  JCAP {\bf 1011}, 026 (2010).
  [arXiv:1008.2641 [astro-ph.CO]].

\bibitem{Riotto:2010nh}
  A.~Riotto, M.~S.~Sloth,
  ``Strongly Scale-dependent Non-Gaussianity,''
  Phys.\ Rev.\  {\bf D83}, 041301 (2011).
  [arXiv:1009.3020 [astro-ph.CO]].

\bibitem{Huang:2011di}
  Q.~-G.~Huang,
  ``Spectral index and running of $g_{NL}$ from an isocurvature scalar field,''
  JCAP {\bf 1104}, 010 (2011).
  [arXiv:1102.4686 [astro-ph.CO]].


\bibitem{Byrnes:2011gh}
  C.~T.~Byrnes, K.~Enqvist, S.~Nurmi and T.~Takahashi,
  ``Strongly scale-dependent polyspectra from curvaton self-interactions,''
  arXiv:1108.2708 [astro-ph.CO].
  
  

\bibitem{Dvali:1998pa}
  G.~R.~Dvali and S.~H.~H.~Tye,
  ``Brane inflation,''
  Phys.\ Lett.\  B {\bf 450}, 72 (1999)
  [arXiv:hep-ph/9812483].


\bibitem{Kachru:2003aw}
  S.~Kachru, R.~Kallosh, A.~D.~Linde and S.~P.~Trivedi,
  ``De Sitter vacua in string theory,''
  Phys.\ Rev.\  D {\bf 68}, 046005 (2003)
  [arXiv:hep-th/0301240].


\bibitem{Langlois:2008wt}
  D.~Langlois, S.~Renaux-Petel, D.~A.~Steer, T.~Tanaka,
  ``Primordial fluctuations and non-Gaussianities in multi-field DBI inflation,''
  Phys.\ Rev.\ Lett.\  {\bf 101}, 061301 (2008).
  [arXiv:0804.3139 [hep-th]].


\bibitem{Mizuno:2009cv}
  S.~Mizuno, F.~Arroja, K.~Koyama, T.~Tanaka,
  ``Lorentz boost and non-Gaussianity in multi-field DBI-inflation,''
  Phys.\ Rev.\  {\bf D80}, 023530 (2009).
  [arXiv:0905.4557 [hep-th]].

\bibitem{Gao:2009at}
    X.~Gao, M.~Li and C.~Lin,
  ``Primordial Non-Gaussianities from the Trispectra in Multiple Field
  Inflationary Models,''
  JCAP {\bf 0911}, 007 (2009)
  [arXiv:0906.1345 [astro-ph.CO]].
  


\bibitem{Starobinsky:1994bd}
  A.~A.~Starobinsky and J.~Yokoyama,
  ``Equilibrium state of a selfinteracting scalar field in the De Sitter
  background,''
  Phys.\ Rev.\  D {\bf 50}, 6357 (1994)
  [arXiv:astro-ph/9407016].




\bibitem{Starobinsky:1986fxa}
  A.~A.~Starobinsky,
  ``Multicomponent de Sitter (Inflationary) Stages and the Generation of
  Perturbations,''
  JETP Lett.\  {\bf 42} (1985) 152;\\  
  M.~Sasaki and E.~D.~Stewart,
  ``A General Analytic Formula For The Spectral Index Of The Density
  Perturbations Produced During Inflation,''
  Prog.\ Theor.\ Phys.\  {\bf 95}, 71 (1996)
  [arXiv:astro-ph/9507001];\\ 
  M.~Sasaki and T.~Tanaka,
  ``Super-horizon scale dynamics of multi-scalar inflation,''
  Prog.\ Theor.\ Phys.\  {\bf 99}, 763 (1998)
  [arXiv:gr-qc/9801017];\\
    D.~H.~Lyth, K.~A.~Malik and M.~Sasaki,
  ``A general proof of the conservation of the curvature perturbation,''
  JCAP {\bf 0505}, 004 (2005)
  [arXiv:astro-ph/0411220];\\  
  D.~H.~Lyth and Y.~Rodriguez,
  ``The inflationary prediction for primordial non-gaussianity,''
  Phys.\ Rev.\ Lett.\  {\bf 95}, 121302 (2005)
  [arXiv:astro-ph/0504045].
  

\bibitem{Komatsu:2010fb}
  E.~Komatsu {\it et al.}  [WMAP Collaboration],
  ``Seven-Year Wilkinson Microwave Anisotropy Probe (WMAP) Observations:
  Cosmological Interpretation,''
  Astrophys.\ J.\ Suppl.\  {\bf 192}, 18 (2011)
  [arXiv:1001.4538 [astro-ph.CO]].
  

\end{thebibliography}
\end{document}